\documentclass[letter,longauth,traditabstract]{aa} % for the long lists of affiliations 
\usepackage{graphicx}
\begin{document}
   \title{The $Herschel$-SPIRE submillimetre spectrum of Mars\thanks{$Herschel$\ 
          is an ESA space observatory with science instruments provided
          by European-led Principal Investigator consortia and
          with participation from NASA.}}
   
  \author{B.\ M.\ Swinyard\inst{1} \and
           P. Hartogh	\inst{2}\and
S. Sidher 	\inst{1}\and
T. Fulton	\inst{3}\and
E. Lellouch 	\inst{5}\and
C. Jarchow 	\inst{2}\and
M.J. Griffin 	\inst{4}\and
R. Moreno 	\inst{5}\and
H. Sagawa 	\inst{2}\and
G. Portyankina 	\inst{6}\and
M. Blecka 	\inst{7}\and
M. Banaszkiewicz 	\inst{7}\and
D. Bockelee-Morvan,  	\inst{5}\and
J. Crovisier 	\inst{5}\and
T. Encrenaz 	\inst{8}\and
M. Kueppers 	\inst{9}\and
L. Lara 	\inst{10}\and
D. Lis 	\inst{11}\and
A. Medvedev 	\inst{2}\and
M. Rengel 	\inst{2}\and
S. Szutowicz	\inst{7}\and
B. Vandenbussche 	\inst{12}\and
F. Bensch 	\inst{13}\and
E. Bergin 	\inst{14}\and
F. Billebaud 	\inst{15}\and
N. Biver 	\inst{5}\and
G. Blake 	\inst{11}\and
J. Blommaert	\inst{12}\and
M. de Val-Borro 	\inst{2}\and
J. Cernicharo 	\inst{16}\and
T. Cavalie 	\inst{2}\and
R. Courtin 	\inst{5}\and
G. Davis 	\inst{17}\and
L. Decin 	\inst{12}\and
P. Encrenaz 	\inst{8}\and
T. de Graauw 	\inst{18}\and
E. Jehin 	\inst{19}\and
M. Kidger 	\inst{9}\and
S. Leeks	\inst{1}\and
G. Orton 	\inst{20}\and
D. Naylor 	\inst{21}\and
R. Schieder 	\inst{22}\and
D. Stam 	\inst{23}\and
N. Thomas 	\inst{24}\and
E. Verdugo 	\inst{9}\and
C. Waelkens 	\inst{12}\and
H. Walker 	\inst{1}}

\institute{STFC Rutherford Appleton Laboratory, Harwell Innovation Campus, Didcot, OX11 0QX, U.K.\and
Max Planck Institute for Solar System Research, Katlenburg-Lindau D-37191, Germany\and
Bluesky Spectroscopy, Lethbridge, Canada\and
School of Physics and Astronomy, Cardiff University, Cardiff, UK\and
LESIA, Observatoire de Paris, France\and
Physikalisches Institut, University of Bern, Switzerland\and
Space Research Centre, Polish Academy of Sciences, Warsaw, Poland\and
LERMA, Observatoire de Paris, France\and
European Space Astronomy Centre, European Space Agency, Madrid, Spain \and
Instituto de Astrofisica de Andalucia, Granada, Spain \and
California Institute of Technology, Pasadena, USA\and
Instituut for Sterrenkunde, Katholieke Universiteit Leuven, Belgium \and
Argelander Institute for Astronomy, University of Bonn, Germany \and
Astronomy Department, University of Michigan, USA \and
Université de Bordeaux, Laboratoire d'Astrophysique de Bordeaux, France \and
Laboratory of Molecular Astrophysics, CAB-CSIC. INTA, Madrid, Spain\and
Joint Astronomy Center, Hilo, USA\and
Joint ALMA Observatory, Chile\and
Institute d'Astrophysique et de Geophysique, Université de Liège, Belgium\and
Jet Propulsion Laboratory, California Institute of Technology, Pasadena, USA\and
Department of Physics and Astronomy, University of Lethbridge, Canada\and
1st Physics Institute, University of Cologne, Germany\and
SRON, Netherlands Institute for Space Research, Netherlands \and
Physikalisches Institut, University of Bern, Switzerland}

   \date{Received 15 April 2010; accepted 20th May 2010}
   
% \abstract{}{}{}{}{} 
% 5 {} token are mandatory
 
  \abstract{We have obtained the first continuous disk averaged spectrum of Mars from 450 to 1550 Ghz using the \it{Herschel}\rm-SPIRE
Fourier Transform Spectrometer. The spectrum was obtained at a constant resolution of 1.4 GHz across the whole band.  The flux from the planet is such that the instrument was operated in``bright
source'' mode to prevent saturation of the detectors.    This was the first successful use of this mode and in this work we describe the method used for observing Mars together with a detailed discussion of the data reduction techniques required to calibrate the spectrum. We discuss the calibration accuracy obtained and describe the first comparison with surface and atmospheric models. In addition to a direct photometric measurement of the planet the spectrum contains the characteristic transitions of $^{12}$CO
from J 5-4 to J 13-12 as well as numerous H$_2$O transitions. Together these allow the comparison to global atmospheric models allowing the mean mixing ratios of water and $^{12}$CO to be investigated.  We find that it is possible to match the observed depth of the absorption features in the spectrum with a fixed water mixing ratio of $1\times10^{-4}$ and a $^{12}$CO mixing ratio of $9\times10^{-4}$.}
  

\maketitle

%________________________________________________________________

\section{Introduction}
Water vapour plays a key role in the Martian atmospheric chemistry and physics. As source of hydrogen radicals it has an important impact on the ozone (Lefevre et al., \cite{le04}) concentration, the stability of the Martian atmosphere (Parkinson and Hunten, \cite{pa72}) and plays an important role in the comparative planetology context (e.g. Hartogh et al. \cite{ha04}, \cite{ha10}). The water cycle and the associated variable hygropause (Clancy \cite{cl96}, Encrenaz \cite{en01}) with a strong asymmetry of its distribution at perihelion and aphelion is one of the principal problems of the present Martian climate. If confirmed, the asymmetry is expected to drive a strong unbalance in the meridional transport. A better observational characterization will be a strong constraint for general circulation models of the Martian atmosphere and particular its meridional transport (e.g. Medvedev and Hartogh, \cite{me07}). Since its first detection in the near infrared (Spinrad et al, \cite{sp63}) water vapour in the Martian atmosphere has been extensively studied by ground-based and space borne observations (e.g. Sprague, et al. \cite{sp96}, Burgdorf, et al. \cite{bu00}, Gurwell, et al. \cite{gu00}, Smith \cite{sm02};  Biver, et al. \cite{bi05}, Fouchet, \cite{fo07}, Encrenaz et al, \cite{en08} and Smith et al, \cite{sm09}). Compared to water, there is only a sparse database for carbon monoxide at Mars. Space borne (Encrenaz, \cite{en06}, and Smith, \cite{sm09}) and ground-based (e.g. Lellouch et al, \cite{ll91}, Krasnopolsky, \cite{kr07}) show an average abundance of 700 ppm.
One of the major goals of the \it{Herschel}\rm~is to trace the origin, evolution and chemistry of water in the Solar system (Hartogh et al \cite{ha09}) using all three \it{Herschel}\rm~instruments, including SPIRE, to take spectra of a variety of Solar system objects including Mars.  Mars was thought to be too bright to be observed using the SPIRE Fourier-transform spectrometer (FTS) as it was anticipated that its bolometric detectors would saturate and provide no useful information.  However, as described below, we have devised a new mode of operation for the instrument that prevents saturation and which has allowed us to use \it{Herschel}\rm-SPIRE to take the first continuous spectrum in the submillimetre of the Martian atmosphere and provide the first complete set of water vapour and carbon monoxide spectra in the 450 to 1550 GHz range.  Given the novelty of this observing mode and the difficulties encountered in reducing and calibrating the data, this paper describes in some detail the observing mode, data reduction and calibration before briefly describing an initial Martian atmospheric model used to validate our calibration approach.  
\begin{figure*}
   \centering
		\includegraphics[angle=0,height=100mm]{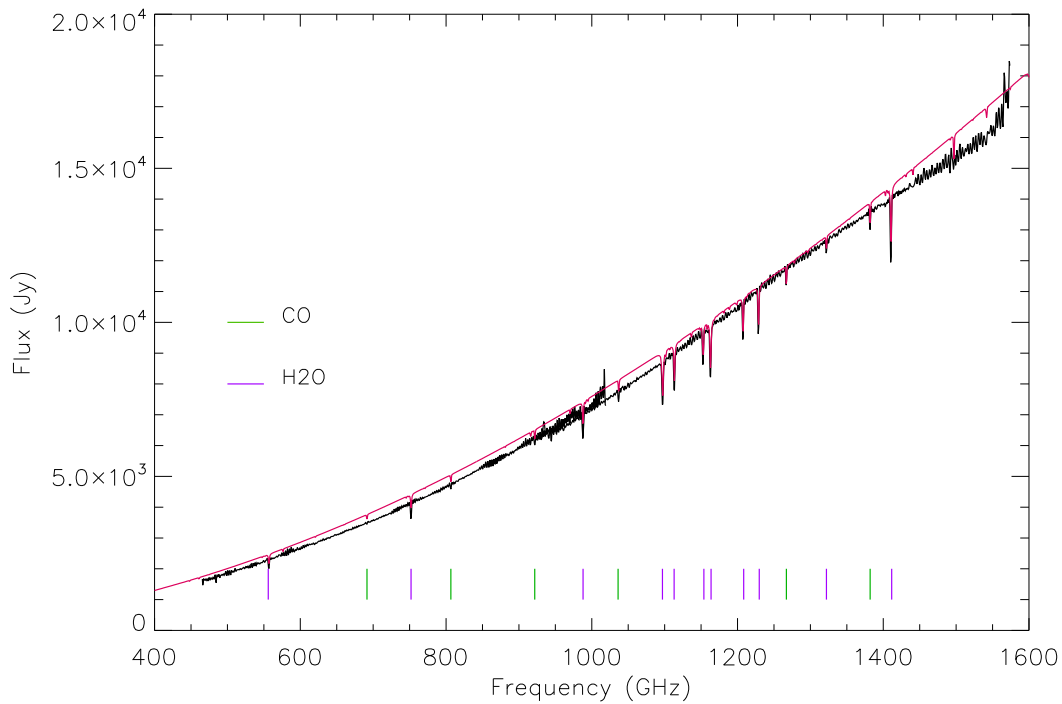}
	\caption{The observed submillimetre spectrum of Mars reduced and calibrated as described in the text compared to a model spectrum generated using the RADTRANS code (Irwin \cite{ir09}).  The positions of the detected H$_2$O and CO lines are indicated.}
	\label{spect}
\end{figure*}
\section{Observations and observing Mode}
\label{observe}
Mars was observed using the central pixels on the SPIRE spectrometer on \it{Herschel}\rm~ operational day 176,  2009 November 6.  See Griffin et al \cite{gr10} for an explanation of the SPIRE instrument and its observing modes.  The observation period commenced at 19:36:16 and finished at 20:20:14 UT: a total of 1332 seconds were spent on the spectrum.  During the observation the target was tracked by the \it{Herschel}\rm~ satellite in non-sidereal pointing mode.  The target was at RA 8h 51m 37.3s and Dec  19$^{o}$15'40.5'' at the start of the observation.  The spectrum was taken as part of the functional and calibration set up observations used to establish the optimum operating conditions for the instrument.  At this time in the mission the notionally 2\% emissivity spectrometer calibration source (SCAL-2) was set to 15 K to assess its efficacy in ``nulling'' the flux from the telescope and thus providing an accurate reference signal against which the source spectrum is measured.  Subsequent observations showed that use of the SCAL is not necessary and, in fact, adds photon noise to the spectrum without adding to the photometric accuracy.
The submillimetre flux density of Mars is extremely high in the context of observing with the SPIRE instrument which was designed to cope with flux densities up to a few hundred Jy rather than the approximately 10000 Jy from Mars. Under nominal detector bias and sensitivity conditions the power from Mars would saturate the detectors and conditioning electronics and no useful data would result.  However, the method of biasing and reading out the bolometers is to use sinusoidal AC bias voltage with a 160 Hz frequency.  The signals from the bolometers are then converted to a DC voltage through a lock-in amplifier that derives its frequency reference from the bias supply electronics.  The long cable (5 m) between the warm electronics and the bolometers in combination with the resistance of the bolometers induces a phase error between the AC bolometer voltage at the input to the lock-in amplifier and the frequency reference.  In normal operations this is removed by calibrating this phase error and maximising the bolometer signal whilst observing ``dark sky''.   However, when observing very bright objects it is desirable to reduce both the inherent sensitivity of the bolometers and the overall electronics gain.  No gain control was built into the SPIRE conditioning electronics so instead we have found that we can vary the gain by deliberately de-phasing the reference and bias waveforms.  The variation in gain between a square wave demodulation and a sine wave is proportional to the cosine of the phase error between them.   In this way then we have a sensitive electronics gain control that can be used to attempt to avoid saturation when observing ``bright'' sources.
%\begin{figure}
%\includegraphics[angle=0,width=70mm]{uranus_pcal_slwc3_temp.eps}
%\caption{Bolometer temperature versus time for the central detector on the SLW array during a PCAL operation.  This example is taken in nominal operating %mode following the observation of Uranus on 2009 December 12.}
%\label{figure1}
%\end{figure}

%\begin{figure}
%\includegraphics[angle=0,width=70mm]{mars_pcal_slwc3_temp.eps}
%\caption{As figure 1 but here taken in ``bright source'' operating mode following the observation of Mars on 2009 November 6.}
%\label{figure2}
%\end{figure}

\section{Data processing and calibration}
For the Mars observation we set the out of phase condition between the reference and bolometer bias to give a notional gain of  about 0.4 for SSW and about 0.28 for SLW compared to the gain when observing dark sky.  In practice, however, the electronics gain is only one part of the overall change in response of the detector system when observing a very bright source such as Mars.  We find that even after applying the known variation in electronics gain the derived flux density of Mars is still much lower than that expected from any model continuum.   During all observations with the SPIRE spectrometer the PCAL source (see Griffin et al \cite{gr10} and Swinyard et al \cite{sw10}) is operated to give a fixed and repeatable power level onto the detectors.  Inspection of the variation in recorded bolometer temperature between the PCAL flash during a reference observation of Uranus and that during the Mars observation after correction for the electronics gain variation shows that the overall response of the system was a factor of 1.5 lower during the Mars observation for the central detector in SLW.  The signal to noise and signal stability for the PCAL data in SSW were not sufficient to allow any estimate of the variation in response and a different approach was adopted here (see below).  
In order to use this information for the Mars data we have processed the interferograms through an alternative processing scheme compared to the standard one described in Fulton et al (\cite{fu08}).  Here we use an observation of Uranus taken on \it{Herschel}\rm~ operational day 209 (2009 December 8) and a model by Moreno (\cite{mo98}, \cite{mo10}) to provide the calibration reference.  We convert the averaged voltage interferograms from the detectors into net bolometer temperature for the dark sky, Mars and Uranus observations separately.  The Uranus and associated dark sky were transformed into spectral space, the dark sky spectra subtracted from the Uranus observation and the net spectrum divided by the Uranus continuum model to provide the calibration relative spectral response function (RSRF).  For the Mars observation the SLW net temperature interferograms were first multiplied by the estimated response variation deduced from the PCAL flashes (x1.5) and then transformed into the spectral domain .  The dark sky observation for Mars was taken with nominal phase settings and no response factor was applied before transformation.  The dark sky was subtracted from the Mars spectrum and the net spectrum divided by the RSRF.  The same procedure was followed for the SSW data but initially with no gain factor applied to the Mars interferogram.  Inspection of the result in comparison to a model (described in Sect. 4) showed a good agreement between the SLW spectrum and the model but a significant variation with the SSW spectrum.  We then applied a variable scaling factor to the SSW Mars interferogram until the recovered flux density in the overlapping spectral region between SSW and SLW came into agreement.  The scaling factor finally applied was 1.35, in reasonable accord with that used for SLW.  
\begin{table}[t]
\caption[]{Principal line identifications in the Mars FTS spectrum. The line to continuum ratios for  H$_2$O lines 11 and 12 are shown in Fig. 2 and for CO line 15 in Fig. 3. $^{*}$This line is blended with 11}
\begin{flushleft}
\begin{tabular}{llll}
\hline\noalign{\smallskip}
Line Number & Species   & Transition & Frequency (GHz)   \\
\noalign{\smallskip}
\hline\noalign{\smallskip}
1 & Ortho H$_2$O  & 1$_{10}$-1$_{01}$ & 556.94    \\
2 & CO            & J=6-5 & 691.491            \\
3 & Para H$_2$O   & 2$_{11}$-2$_{02}$ & 752.03    \\
4 & CO            & J=7-6 & 806.65            \\
5 & CO            & J=8-7 & 921.8             \\
6 & Para H$_2$O   & 2$_{02}$-1$_{11}$ & 987.93   \\
7 & CO            & J=9-8 & 1036.91            \\
8 & Ortho H$_2$O  & 3$_{12}$-3$_{03}$ & 1097.37    \\
9 & Para H$_2$O   & 1$_{11}$-0$_{00}$ & 1113.34    \\
10 & CO            & J=10-9 & 1152.01$^{*}$             \\
11 & Ortho H$_2$O  & 3$_{12}$-2$_{21}$ & 1153.13    \\
12 & Ortho H$_2$O  & 3$_{21}$-3$_{12}$ & 1162.91   \\
13 & Para H$_2$O   & 4$_{22}$-4$_{13}$ & 1207.64    \\
14 & Para H$_2$O   & 2$_{20}$-2$_{11}$ & 1228.79    \\
15 & CO            & J=11-10 & 1267.01            \\
16 & H$_2$O   & 6$_{25}$-5$_{32}$ & 1322.06                   \\
17 & CO            & J=12-11 & 1382.0            \\
18 & H$_2$O   & 5$_{23}$-5$_{14}$ & 1410.6                   \\
\hline
\end{tabular}
\end{flushleft}
\end{table}
The final result is shown in Fig. 1 together with the model spectrum.  We can see that there is now good agreement in the flux density across the SPIRE range but some distortion of the continuum spectrum in the SSW range as well as a small scaling difference between the overall continuum and the measured spectrum amounting to between 5 and 10\% across waveband.  Inspection of the voltage timelines shows that, even with the reduced electronics gain there is still a degree of saturation in SSW when observing Mars and this causes the low order part of the transformation to be distorted.  The level of agreement is satisfactory and, importantly for the overall interpretation of the spectrum, is demonstrated to be only subject to scaling and not offset errors. The absolute calibration accuracy using this method can be estimated from the uncertainties in the ratio of the PCAL flashes and to first order are around 20\% for the SLW detector - similar to those quoted for the nominal operational mode of the SPIRE FTS (Swinyard et al \cite{sw10}).  However for the SSW we do not have an independent measurement of the uncertainty and, given that there are also distortions evident in the continuum spectrum, we can assign an accuracy of no better than 30\% at present.  New observational and calibration techniques for bright sources are under development and we expect this to improve.
\section{Atmospheric model of Mars}
The observation was modelled using the RADTRANS line-by-line radiative transfer code (Irwin \cite{ir09}) and the HITRAN spectral database. We prescribed an input atmospheric temperature profile and a surface brightness temperature by using outputs from the European Martian Climate Database (EMCD, v4.1). The EMCD is a database describing the climate and environment of the Martian atmosphere, derived from martian general circulation models (Forget \cite{for99}, Lewis et al. \cite{lew99}).
Essentially, for the appearance and season (L$_s$, sol) of Mars relevant to our observation, we partitioned the visible martian disk in a fine, regular grid (typically 100 points across a martian diameter). The EMCD database was then used to provide the local surface pressure and the vertical profiles of temperature in the atmosphere and in the sub-surface. The surface pressure and the temperature-pressure-altitude profiles were averaged over the disk, providing the input thermal profile needed for the radiative transfer calculation. For calculating the surface brightness temperature, radiative transfer was performed within the surface, using the online tool of Lellouch et al.\ (\cite{ll08}): a surface brightness temperature of 220 K was derived.  Finally, the derived spectrum was convolved with a sinc function of FWHM 0.048cm$^{-1}$ and averaged over the disk.  For the CO lines we tried mixing ratios between 4.5 and 18$\times 10^{-4}$ and for the H$_2$O lines we tried mixing ratios between 0.5$\times 10^{-4}$ and 2$\times 10^{-4}$.  The disk-averaged flux density was then finally calculated given the Mars solid angle of 1.25$\times 10^{-10}$ for the observation date of 2009 November 6.

The full spectrum model generated with fixed mixing ratios of 9$\times 10^{-4}$ for CO and 1$\times 10^{-4}$ for H$_2$O is plotted together with the calibrated observed spectrum in Fig. 1 and the identified lines are given in Table 1.  Note that the CO 4-3 and 4-5 lines were not clearly detected in the spectrum as these regions suffered from instrumentally induced spectral fringes which masked the absorption features.  The level of agreement between the model continuum level and the observed spectrum is good except in the SSW range where some distortion is found for reasons described above.  A more detailed view of the H$_2$O line to continuum ratios is shown in Fig. 2.  We have taken three model cases here to show the variation of line depth and width with constant mixing ratios of 2$\times 10^{-4}$ (purple), 1$\times 10^{-4}$ (green) and 0.5$\times 10^{-4}$ (red).  It is evident that the green line shows the best fit to the line depths.  However, the line wings in the model appear broader than in the observation and further more detailed modelling in conjunction with higher spectral resolution $Herschel$-HIFI observations will be required to fully match the spectrum.  In Fig. 3 we show a close up of the spectrum around the CO J=11-10 line showing the effect of CO mixing ratios of 18$\times 10^{-4}$ (purple), 9$\times 10^{-4}$ (green) and 4.5$\times 10^{-4}$ (red).  Over all CO lines 9$\times 10^{-4}$ was found to be the best fit.

\begin{figure}
\includegraphics[angle=0,width=80mm]{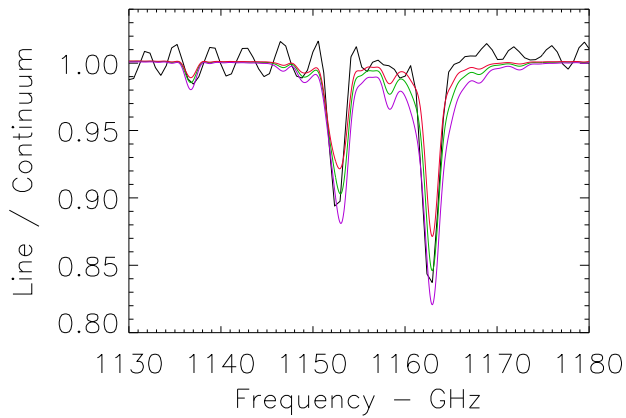}

\caption{Close up of the measured and model spectrum around the Ortho 3$_{12}$-2$_{21}$ and 3$_{21}$-3$_{12}$ water lines showing the effect of varying the mixing ratio (see text)}
\label{close}
\end{figure}
\begin{figure}
\includegraphics[angle=0,width=80mm]{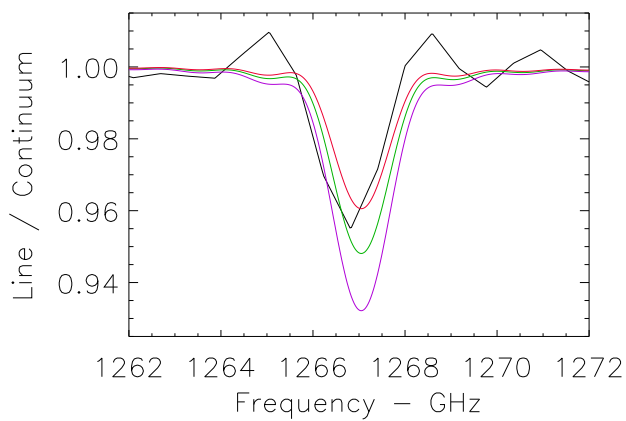}

\caption{Close up of the measured and model spectrum around the CO J=11-10 line showing the effect of varying the mixing ratio (see text)}
\label{CO}
\end{figure}
\section{Conclusions}
\label{conclusions}
This paper has demonstrated the utility of the SPIRE FTS in making efficient broad band spectral observations of very bright astronomical sources and the ability to provide accurately calibrated spectra when SPIRE is operated in its ``bright source'' mode.  This was not expected to be the case before launch and it represents a significant enhancement to the observation capabilities of \it{Herschel}\rm.  We have taken the first continuous submillimetre spectrum of the planet Mars and shown the ability to measure all water and carbon monoxide lines between 600 and 1550 GHz with good photometric accuracy in a single short observation.  Our preliminary analysis of the spectrum shows that a model with abundances of 900 ppm for CO and 100 ppm for H$_2$O best fits the observed line depths.  Both these values are in line with previously established figures.  A more sophiscated modeling approach is clearly needed to make the best use of these and future data sets that will be obtained during the \it{Herschel}\rm~ mission and this will be the subject of future work.

\begin{acknowledgements}
SPIRE has been developed by a consortium of institutes led by Cardiff Univ. (UK) and including Univ. Lethbridge (Canada); NAOC (China); CEA, LAM (France); IFSI, Univ. Padua (Italy); IAC (Spain); Stockholm Observatory (Sweden); Imperial College London, RAL, UCL-MSSL, UKATC, Univ. Sussex (UK); Caltech, JPL, NHSC, Univ. Colorado (USA). This development has been supported by national funding agencies: CSA (Canada); NAOC (China); CEA, CNES, CNRS (France); ASI (Italy); MCINN (Spain); SNSB (Sweden); STFC (UK); and NASA (USA)     
We acknowledge the generous support in running the RADTRANS code provided by Pat Irwin of the Oxford Planetary Group, Atmospheric, Oceanic and Planetary Physics, Clarendon Laboratory.
\end{acknowledgements}

\end{document}